\documentclass[prl,aps,showpacs,twocolumn,unsortedaddress]{revtex4}
\pdfoutput=1
\usepackage{graphics, bm}
\usepackage{psfrag}
\usepackage{amsmath}
\usepackage{amssymb}
\usepackage{epsfig}
\usepackage{subfigure}
\newcommand{\eps}    {\epsilon}
\newcommand{\beq}    {\begin{equation}}
\newcommand{\enq}    {\end{equation}}

\newcommand{\rr}     {{\bf r}}

\newcommand{\df}     {\equiv}

\newcommand{\dngm}   {\Delta n_{\rm tg}}
\begin{document}

\title{Signatures of Klein tunneling in disordered graphene p-n-p junctions}

\author{E.~Rossi$^1$, J.~H.~Bardarson$^{2}$, P.~W.~Brouwer$^{2,3}$ and S.~Das~Sarma$^1$}
\affiliation{$^1$Condensed Matter Theory Center, Department of Physics, University of Maryland, College Park, Maryland 20742-4111}
\affiliation{$^2$Laboratory of Atomic and Solid State Physics,
Cornell University, Ithaca, NY 14853-2501}
\affiliation{$^3$Dahlem Center for Complex Quantum Systems and Institut f\"ur
Theoretische Physik, Freie Universit\"at Berlin, Arnimallee 14, 14195 Berlin,
Germany}

\date{\today}


\begin{abstract}
We present a method for obtaining quantum transport properties in graphene that uniquely combines three crucial features:
microscopic treatment of charge disorder, fully quantum mechanical analysis of transport, and the ability to model experimentally
relevant system sizes. As a pertinent application we study the disorder dependence of Klein tunneling dominated transport in {\em
p-n-p} junctions. Both the resistance and the Fano factor show broad resonance peaks due to the presence of quasi bound states.
This
feature is washed out by the disorder when the mean free path becomes of the order of the distance between the two {\em p-n}
interfaces. 

\end{abstract}


\maketitle


A realistic modeling of quantum transport in graphene~\cite{novoselov2004} has two main
requirements: microscopic treatment of disorder and a fully quantum mechanical calculation of transport, taking into account
the finite (but large with respect to the lattice constant) size of the samples. This is especially important close to the Dirac
point~\cite{neto2009}, the charge neutrality point at which transport is dominated by evanescent modes and conventional analytical methods fail, and for interference effects.
In this letter we report on the first such calculation in a study of the resistance oscillations in a disordered graphene
{\em p-n-p} junction which incorporates both realistic Coulomb
disorder and quantum mechanical transport. 

The physics of the {\em p-n-p} junction is governed by ``Klein
tunneling'', an effect first discovered in the context of a
relativistic particle tunneling through a potential barrier with a
height comparable to its rest energy~\cite{dombey1999}. 
Since the occurrence of relativistic Klein tunneling is a crucial
feature distinguishing graphene from ordinary electronic materials, a
clear understanding of how to unambiguously observe Klein tunneling in
realistic graphene samples, where disorder is unavoidably present, is
important. 
In graphene,
transmission through a potential barrier has a pronounced angular
dependence, with perfect transmission for perpendicular incidence and
a quick decrease of transmission probability for finite angle of
incidence~\cite{ando98}. 
As a consequence, electrons can be confined within a single potential
barrier, such as the {\em p-n-p}
junction~\cite{silvestrov2007,bardarson2009}. Resonant tunneling
through those confined states then leads to pronounced oscillations
in the resistance as a function of system
parameters~\cite{silvestrov2007, shytov2008, bardarson2009}.

The existence of this phenomenon relies heavily on the presence of a
well defined interface between the {\em p} and {\em n} regions.  Early
experiments failed to reproduce the predicted resistance
oscillations~\cite{huard2007}, 
presumably because of too much disorder. 
More recent experiments that seek to
minimize the effects of disorder have been more
successful~\cite{gorbachev2008, stander2009, young2009}. 
However a theoretical understanding of whether or to what extent
these experiments are really observing Klein tunneling phenomena and
the issue of the experimental conditions needed to see Klein tunneling
unambiguously have remained open.

There exists strong evidence that disorder in current
experiments is dominated by remote charged impurities~\cite{tan2007,
chen2008} invariably present in the graphene environment. 
Due to the long range character of the Coulomb potential
of impurity and gate charges, all scattering potentials and charge
densities in the graphene sheet vary slowly on the scale of the
lattice constant. This fact leads to two simplifications that make
an accurate modeling of current experimental 
setups~\cite{gorbachev2008, stander2009, young2009} possible.
First, the Thomas-Fermi-Dirac (TFD)
self-consistent density functional method is sufficient
to obtain the ground state carrier density in the presence of 
disorder~\cite{rossi2008,rossi2009}.
Second, intervalley coupling can
be neglected for a smooth scattering potential, so that one can model 
transport using the single valley Dirac Hamiltonian. The transfer 
matrix method of Refs.~\onlinecite{titov2007}~and~\onlinecite{bardarson2007} is 
ideal for this purpose and can be used for experimentally relevant 
system sizes, as opposed to tight binding models which are
generally limited to small system sizes.

\begin{figure}[tb]
 \begin{center}
  \includegraphics[width=8.5cm]{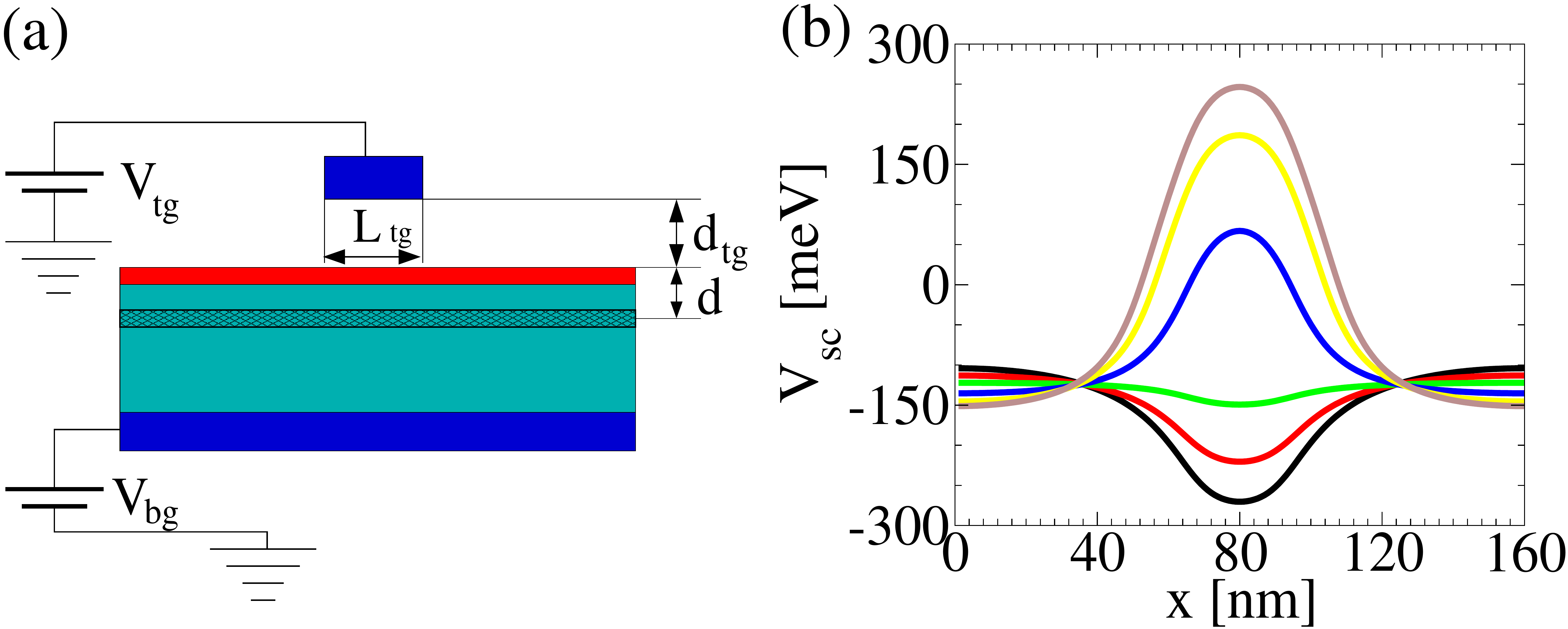}
  \caption{
           (Color online). 
           (a) A schematic of the top gated graphene setup studied.
           (b) Profile of the screened potential $V_{\rm sc}$ in the absence of disorder. The back gate density is fixed at 
           $n_{\rm bg} = 5\times 10^{11} {\rm cm}^{-2}$ and (from bottom to top) $\dngm=-5\times 10^{12} {\rm cm^{-2}}$ 
           to $5\times 10^{12} {\rm cm^{-2}}$ in steps of $2\times 10^{12} {\rm cm^{-2}}$.
           All results in this paper were obtained for square samples of size $W=L=160$~nm assuming
           $L_{\rm tg}=30$~nm, $d_{\rm tg}=10$~nm, $d=1$~nm, $\eps_1=1$ (air), and $\eps_2=4$ (${\rm Si O_2}$).
          } 
  \label{fig:profiles}
 \end{center}
\end{figure} 

Our calculation thus consists of two steps. In the first step the ground state carrier density $n(\rr)$ and the corresponding screened Coulomb
potential $V_{\rm sc}$ is obtained using the TFD approach \cite{rossi2008}. In the second step the resistance and Fano factor, the ratio of shot noise power and electrical current, are calculated using $V_{\rm sc}$ as an input into the fully quantum mechanical
transfer matrix approach~\cite{titov2007, bardarson2007}. We first describe the two steps of our method in more detail and then turn to the results. 

A schematic of the {\em p-n-p} junction under consideration is shown in Fig.~\ref{fig:profiles}a. 
A top gate of length $L_{\rm tg}$ is located a distance $d_{\rm tg}$ above the graphene layer --- in an air-bridge
setup the medium in between is air with dielectric constant $\eps_1 = 1$. A back gate that controls the average carrier density is
separated from the graphene by a SiO$_2$ substrate (dielectric constant $\eps_2 = 4$). The effective
dielectric constant in this setup is $\eps=(\eps_1+\eps_2)/2$. Together, the voltages on these two gates define the junction. Examples of potential profiles are shown in Fig.\ \ref{fig:profiles}b.

A number of charged impurities is trapped in the substrate and just below the graphene layer. We model this by a random distribution 
$C(\rr)$ of impurity charges at a fixed distance $d$ with properties
\begin{equation}
 \langle C(\rr)\rangle = 0; \hspace{0.5cm}
 \langle C(\rr_1) C(\rr_2)\rangle = n_{\rm imp}\delta(\rr_2 - \rr_1);
 \label{eq:C_stat}
\end{equation}
with $n_{\rm imp}$ the average 2D charge impurity density. 

In the TFD approximation, the ground state carrier density $n(\rr)$
in the graphene layer is obtained by minimizing the energy functional
\begin{eqnarray}
  E[n] &=& \int d^2\rr n(\rr) 
           \left[ \frac{2}{3} \hbar v_F |\pi n(\rr)|^{1/2} 
           + V_{\rm sc}(\rr) \right]
  \nonumber \\ && \mbox{}
+ E_{\rm xc}[n]
  \label{eq:en}
\end{eqnarray}
with respect to $n(\rr)$. Here $v_F \approx 10^6\;{\rm m/s}$ is the Fermi 
velocity for graphene, 
\begin{eqnarray}
  V_{\rm sc}(\rr) &=& \hbar v_F r_s \left[V_{\rm d}(\rr) + V_{\rm tg}(\rr)
  + \frac{1}{2}\int d^2 r'\frac{n(\rr')}{|\rr-\rr'|} \right] 
 \nonumber \\ && \mbox{} - \hbar v_F \mu,
 \label{eq:vsc}
\end{eqnarray}
with $r_s\df e^2/\hbar v_F\epsilon$, $V_{\rm d}$ and $V_{\rm tg}$ the
potentials induced by the impurity density 
$C(\rr)$ and the top gate, respectively, and $\mu$ the chemical potential.
$E_{\rm xc}[n]$ is the exchange correlation energy 
(see \cite{rossi2008,polini2008} for details) which we include even though
it gives only minor quantitative corrections to 
the transport properties of the {\em p-n-p} junction. 
The minimization is subject to the constraint $(1/A')\int_{A'} n(\rr) d^2r = n_{\rm bg}$,
where $A'$ is the area of the sample away from the top gate.
The constraint is enforced self-consistently by varying $\mu$.
The electrostatic potential 
$V_{\rm tg}$ is expressed in terms of the top gate charge density 
$n_{\rm tg}$,
\begin{equation}
 V_{\rm tg}(\rr) = \int d^2\rr'\frac{n_{\rm tg}(\rr')}{(|\rr-\rr'|^2 + d_{\rm tg}^2)^{1/2}},
 \label{eq:vtg}
\end{equation}
and is obtained self-consistently, 
for a fixed voltage difference between the top gate and graphene, $\varphi_{\rm tg}$,
by requiring that in the region below the top gate
$
  \Delta n_{\rm tg} \equiv 
  n_{\rm tg}(\rr)-n(\rr) = C_{\rm tg} \varphi_{\rm tg},
$
where $C_{\rm tg}$ is the top gate capacitance. 

The potential $V_{\rm sc}$ defines a scattering problem through the Dirac Hamiltonian
\begin{equation}
  H = v_F \mathbf{p}\cdot \mathbf{\sigma} + V_{\rm sc}(\rr)
\end{equation}
with $\sigma = (\sigma_x, \sigma_y)$ the Pauli matrices. The Schr\"odinger equation $H \psi = 0$ generates a transfer matrix $\mathcal{M}$, which relates
the wavefunction at $x=0$ to the one at $x=L$, $\psi_L = \mathcal{M}\psi_0$. In order to numerically calculate $\mathcal{M}$ we divide the
interval $(0,L)$ into $N$ equal subintervals of length $\delta x = L/N$ and calculate the transfer matrix in each subinterval in the
Born approximation. This gives~\cite{bardarson2007} 
\begin{equation}
  \mathcal{M} = \prod_{n=1}^N 
  e^{- \frac{i}{2} \delta x \partial_y \sigma_z}
  e^{-i u_n \sigma_x}
  e^{- \frac{i}{2} \delta x \partial_y \sigma_z},
\end{equation}
where
\begin{equation}
  \label{eq:Un}
  u_n(y) = \frac{1}{\hbar v_F}
  \int_{(n-1)\delta x}^{n\delta x} dx\, V_{\rm sc}(x,y).
\end{equation}
We then take the limit $N\rightarrow \infty$ in which the Born
approximation becomes exact. From the transfer matrix $\mathcal{M}$ we
calculate the matrix $t$ of transmission amplitudes, which in turn
gives us the two terminal conductance and Fano factor as $G = R^{-1}
= (4e^2/h)
{\rm tr}\, tt^\dagger$ and $F = {\rm tr}\,
[(1-tt^\dagger)\,tt^\dagger]/{\rm tr}\, tt^\dagger$. 

\begin{figure}[tb]
 \begin{center}
  \includegraphics[width=8.5cm]{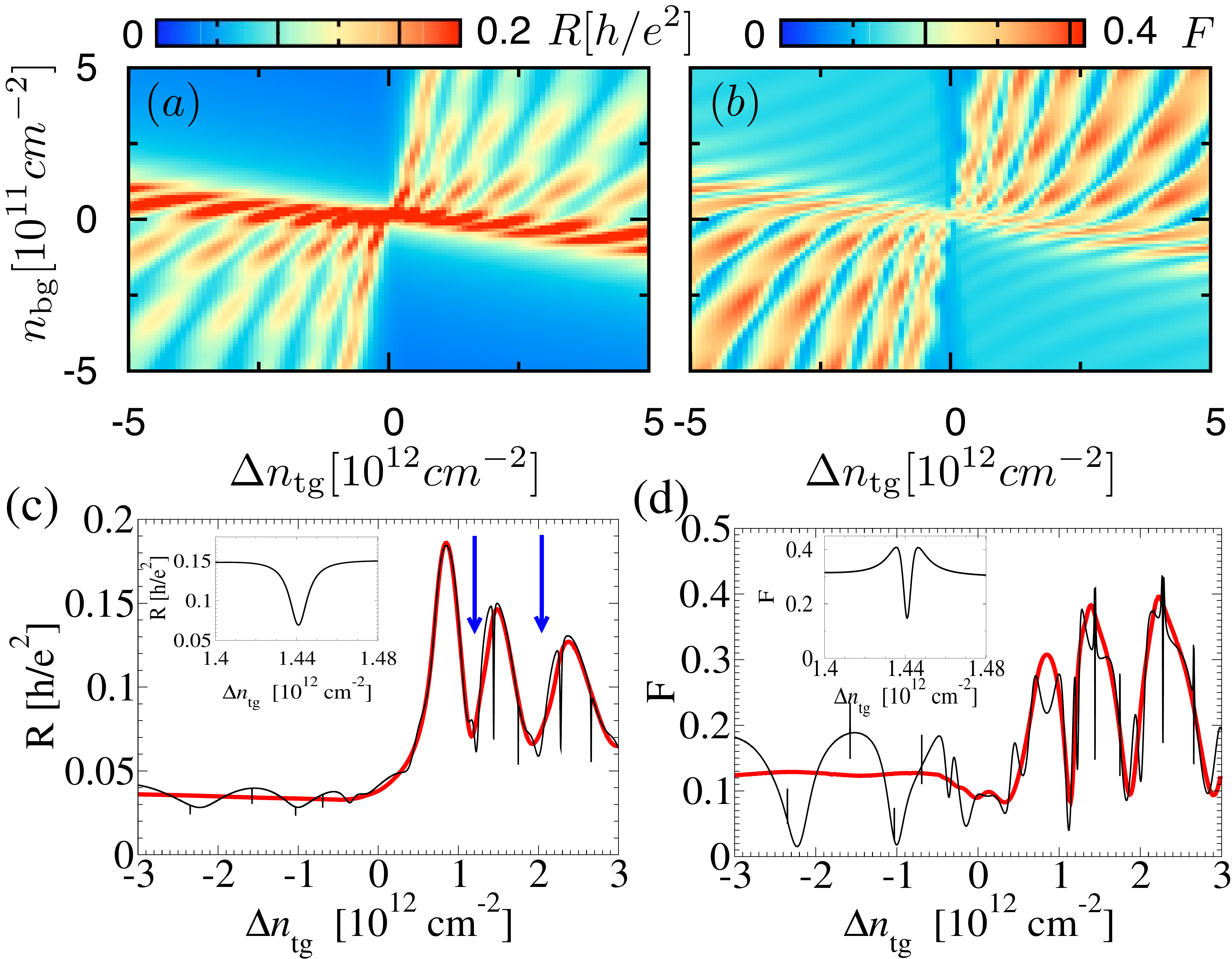}
  \caption{
           (Color online).
           Gate voltage dependence of (a) resistance and (b) Fano factor of a clean device. 
           A single trace of (c) resistance and (d) Fano
           factor at a fixed back gate density $n_{\rm bg}=5\times 10^{11}{\rm cm}^{-2}$. The smooth curves are obtained by averaging over boundary conditions. The insets show a close up of
           one of the narrow resonances obtained for transverse periodic boundary condition.
            }
            \label{fig:clean}
 \end{center}
\end{figure} 
We now turn to our results. We first consider the clean case $n_{\rm
imp} = 0$. By varying the top gate potential, the junction type can be
varied from {\em p-n-p} to {\em p-p-p} (or {\em n-p-n} to {\em n-n-n}
depending on the back gate voltage), as shown in
Fig.~\ref{fig:profiles}b. The dependence of the resistance $R$ and Fano
factor $F$ on the two gate voltages is shown in
Fig.~\ref{fig:clean}a-b. In the {\em p-p-p} ({\em n-n-n}) region of the
upper left (lower right) corner the resistance is small compared to
the resistance of the {\em p-n-p} ({\em n-p-n}) region of the upper
right (lower left) corner. Pronounced oscillations in the resistance are seen
in the {\em p-n-p} and {\em
n-p-n} regions. The behavior of the Fano factor mostly follows that of
the resistance.

The resistance oscillations can be understood as arising from resonances through quasi bound states inside the barrier
created by the two {\em p-n} interfaces~\cite{silvestrov2007, shytov2008, bardarson2009}. The larger the transverse momentum $q_y$, the
larger the incident angle for scattering off the barrier and the tunneling amplitude decreases. The broad oscillations arise
from resonant tunneling of the few modes with smallest $q_y$. Larger $q_y$ give rise to very narrow resonance seen
in Fig.~\ref{fig:clean}c-d, where we show a cross-section of
Fig.~\ref{fig:clean}a-b at a fixed value of the back gate voltage in a higher
resolution. Since the positions and widths of the narrow resonances are sensitive 
to the transverse boundary conditions we also plot the smooth curves obtained by averaging over twisted boundary conditions. The
weak oscillations in the {\em p-p-p} and {\em n-n-n} regions and the narrow resonances are absent after the averaging.
Observation of the narrow resonances will thus require very well defined edges.
The double peak structure of the Fano factor resonances
reflects the increase of the transmission probability from zero
to one and back to zero and the fact that the Fano factor of a perfectly transmitted mode is zero.

In contrast to the narrow resonances the resistance minima for the 
broader oscillations do not depend on boundary conditions and can be estimated by a semiclassical
argument~\cite{silvestrov2007}. One can think of the {\em p-n-p} 
junction as a Fabry-Perot etalon in which
waves scattered off the two {\em p-n} interfaces interfere destructively 
if the WKB phase
\beq
 \theta_{\rm WKB} = -\int_{x_1}^{x_2}V_{\rm sc}(x',y)d x'
 \label{eq:thetawkb} \equiv 0 \ \mbox{(mod $\pi$)},
\enq 
with the turning points $x_i$ defined by the condition
$V_{\rm sc}(x_i,y) = 0$, $i=1,2$. 
The positions of the resistance
minima that follow from this argument are indicated by arrows in 
Fig.~\ref{fig:clean}.

\begin{figure}[tb]
   \begin{center}
    \includegraphics[width=8.5cm]{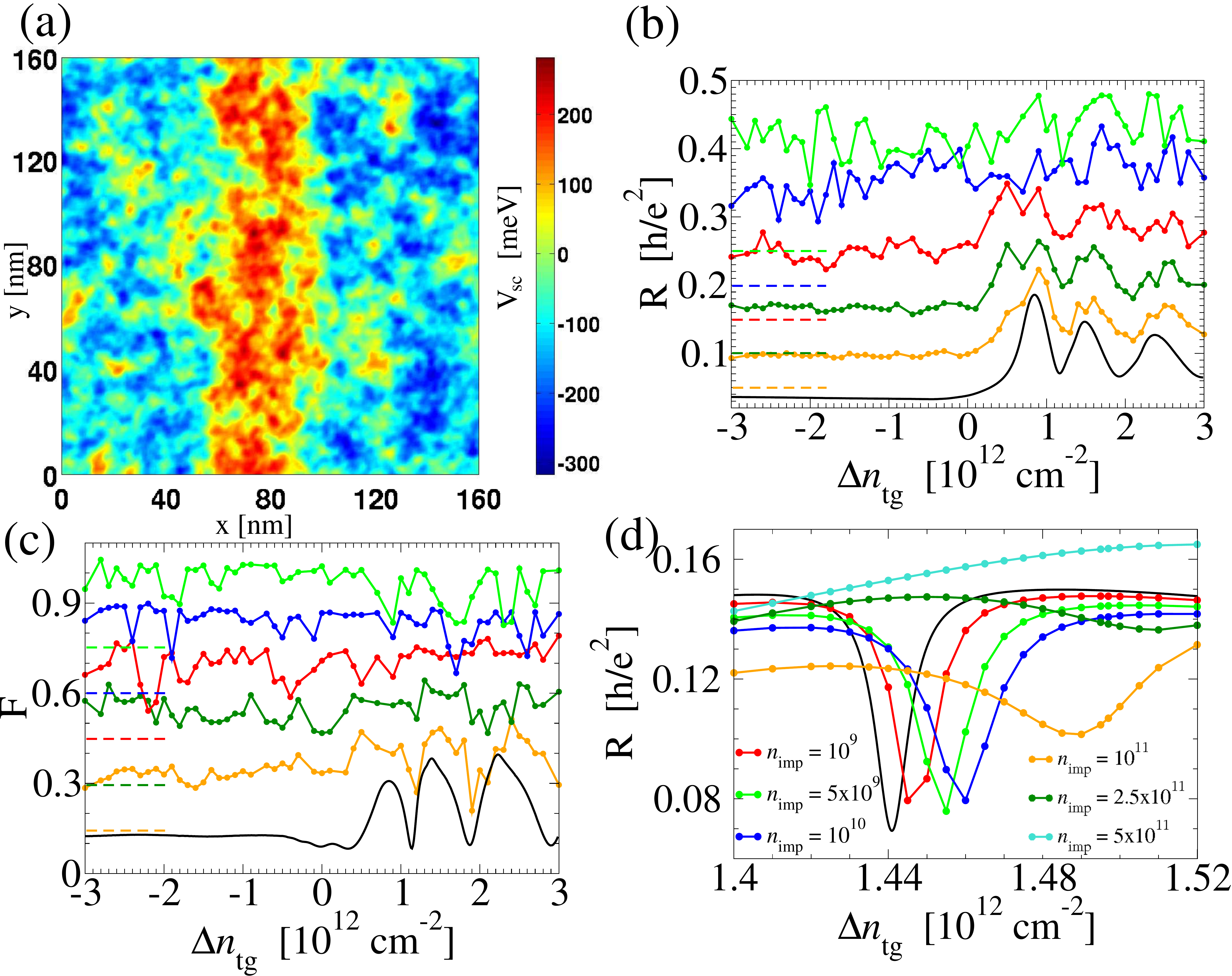}
  \caption{
           (Color online). 
           (a) Density $n(\rr)$ for a disordered junction with 
           $n_{\rm bg}=5\times 10^{11} {\rm cm^{-2}}$, 
           $\Delta n_{\rm tg}=2\times 10^{12} {\rm cm^{-2}}$, and
           $n_{\rm imp}=5\times 10^{11} {\rm cm^{-2}}$. 
           (b) resistance and (c) Fano factor in the presence of disorder (single realization) for a fixed
           back gate density $n_{\rm bg}=5\times 10^{11}{\rm cm}^{-2}$ and several values of the impurity density (from bottom to top $n_{\rm
           imp} = 0, 1, 2.5, 5, 10,$ and $15 \times 10^{11}{\rm cm}^{-2}$). The disordered curves have been shifted for clarity, with
           dashed lines showing their zero. (d) The top gate density dependence of the resistance at a narrow resonance for different values of the
           impurity density.
          } 
  \label{fig:RFsingle}
 \end{center}
\end{figure}

We now consider the effect of disorder. A single realization of the
scattering potential $V_{\rm sc}$ in a disordered {\em p-n-p} junction
is shown in Fig.~\ref{fig:RFsingle}a. For realistic impurity
concentrations, the correlation length of $V_{\rm sc}$ is $\sim 10$~nm
\cite{rossi2008, zhang2009}. While one can still clearly make out the
different $p$ and $n$ regions in the presence of disorder, the
boundary between the two is no longer as sharp, leading to a weaker
confinement of particles inside the barrier. The effects of Klein
tunneling are thus expected to be suppressed.

Figures~\ref{fig:RFsingle}b-c give the transport properties
as a function of top gate voltage for a single disorder realization
and for different impurity densities. Mesoscopic fluctuations due to
the disorder are superimposed on the oscillations from the Klein
tunneling. As the impurity strength increases these fluctuations
become stronger, eventually dominating the signal. The Fano factor is
more sensitive to disorder than the resistance. The effect of disorder
on a narrow resonance is shown in Fig.~\ref{fig:RFsingle}d.

\begin{figure}[tb]
 \begin{center}
  \includegraphics[width=8.5cm]{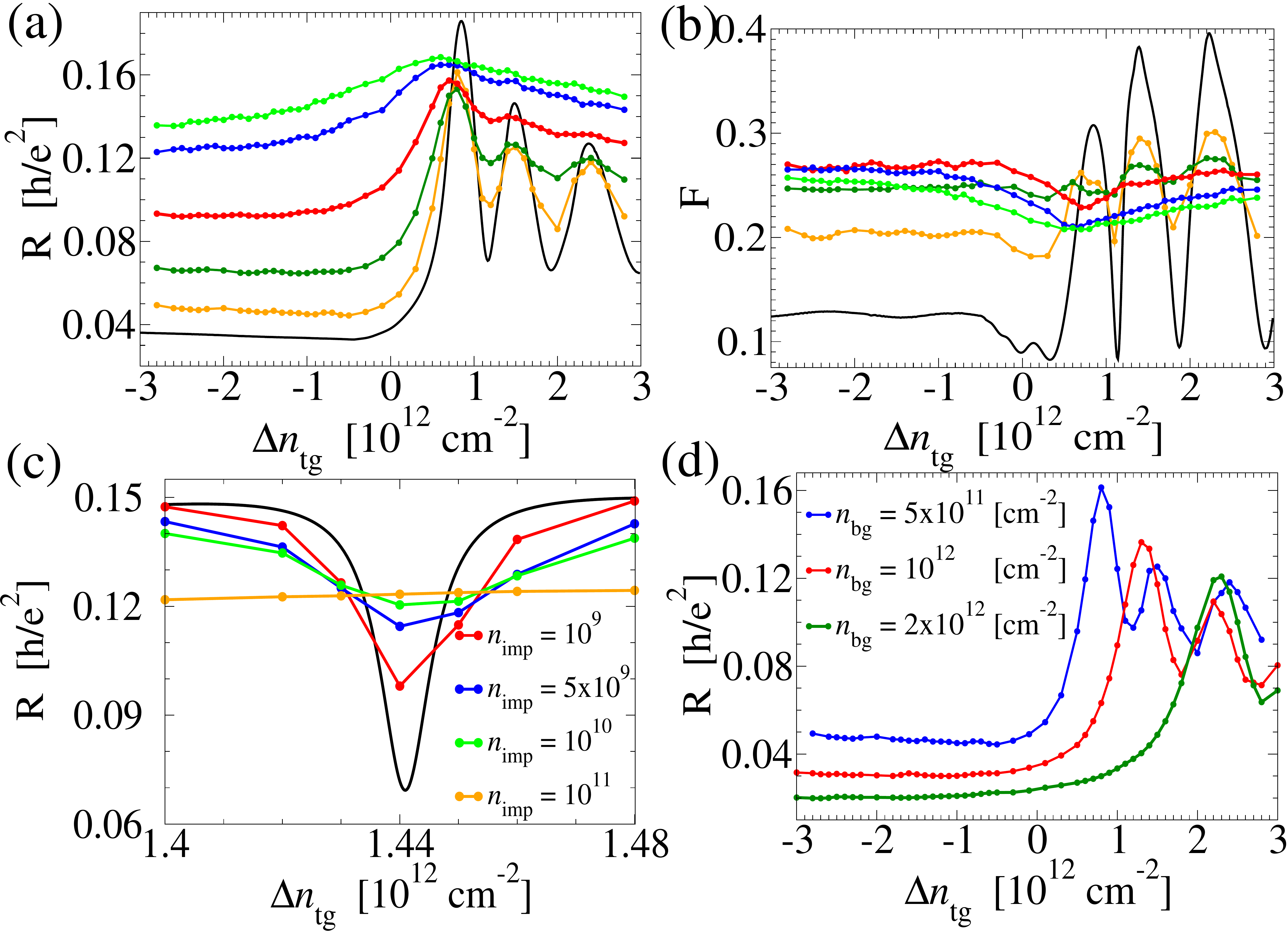}
  \caption{           
           (Color online). 
           Disorder averaged resistance (a) and Fano factor (b) as a function of top gate voltage for different values of impurity
           densities (parameters and color coding the same as in Fig.~\ref{fig:RFsingle}~b. (c) Disorder averaged resistance at a
           narrow resonance for different values of the impurity density. 
           (d) Top-gate density dependence of the disorder averaged resistance
           for different values of back gate density.          
          } 
  \label{fig:RFdisorder}
 \end{center}
\end{figure}

In Fig.~\ref{fig:RFdisorder} we show the disorder averaged resistance and Fano factor, where the average was taken over $10^3$
disorder realizations. We find that the clean limit broad quantum oscillations survive for impurity densities up
to $n_{\rm imp} \sim 10^{12}$ cm$^{-2}$ while the narrow resonances disappear at $n_{\rm imp} \sim 10^{11}$ cm$^{-2}$.
We do emphasize that for weak enough disorder the narrow resonances
will be present, and it is therefore conceivable that they may
actually be observed at low enough temperatures, when the phase
coherence length is much larger than the system size.

As in the case of the single disorder realization, the dependence of the Fano factor on the
disorder strength is similar to that of the resistance.
The dependence of the resistance on the back gate voltage for a fixed impurity density is shown in Fig.~\ref{fig:RFdisorder}d. 
The main effect of the back gate is an overall reduction of the resistance and a shift of the top gate density at which the central
region changes polarity. 

One expects impurity scattering effects to start dominating the
ballistic oscillations when the mean free path $l$ becomes comparable
to the sample width $W$ (for the narrow resonances) or the length of 
the central $n$ region (for the broad oscillations). 
The mean free path $l$ can
be estimated through its relation to the conductivity $\sigma$, $l = (h/2
e^2)(\sigma/k_F)$, where $k_F$ is the
Fermi wavevector. Away from the Dirac point, $\sigma=(2e^2/h)\langle
n\rangle/n_{\rm imp}f(r_s,d)$ 
\cite{nomura2006},
where $f(r_s,d)$ is a known function of $r_s$ and $d$ only
[$f(r_s=0.8,d=1\,{\rm nm})=0.1$] \cite{adam2007}. 
Setting $\langle n\rangle=n_{\rm bg}$ and 
using the impurity densities quoted above, we find that 
the narrow resonances vanish for $l \approx 400$~nm, which is the same order of magnitude as expected.
The impurity density at which the broad oscillations disappear corresponds to a mean free path 
$l \approx 40$~nm. This is roughly the same as the length between the
two {\em p-n} interfaces, see Fig.~\ref{fig:profiles}, and consistent 
with the experimental results of
Refs.~\onlinecite{gorbachev2008} and \onlinecite{stander2009}.

In summary, we have presented a powerful theoretical method that is
generally applicable to transport problems in realistic graphene
samples where the disorder is dominated by charge impurities and
transport properties need to be obtained fully quantum mechanically.
We have applied this method to understand the effects of disorder on
transport through {\em p-n-p} junctions. The crossover from the
ballistic transport governed by Klein tunneling, to the disordered
diffusive transport is found to take place as the mean free path
becomes of the order of the distance between the two {\em p-n}
interfaces
consistent with recent 
experiments~\cite{gorbachev2008, stander2009}.

We thank  S. Adam, E.~H. Hwang, and
in particular  L.~S.~Levitov for discussions.
The numerical calculations have been performed on the University
of Maryland High Performance Computing Cluster (HPCC).
This work is supported by US-ONR and NSF-NRI, by the NSF under
grant no.\ DMR 0705476, and by the Humboldt Foundation. JHB thanks the Dahlem Center at FU Berlin for hospitality.




\end{document}